\documentclass[12pt]{article}

\usepackage{natbib}
\usepackage{amssymb,alltt}
\usepackage{graphicx, framed, enumitem,amsmath, listings}
\usepackage{float}
\usepackage{graphicx}
\usepackage{subcaption}
\usepackage{url}
\usepackage[pdftex,pdfpagelabels,bookmarks,hyperfigures]{hyperref}
\usepackage{color}
\usepackage{rotating}
\usepackage[parfill]{parskip}
\usepackage [autostyle, english = american]{csquotes}
\MakeOuterQuote{"}
\usepackage{epigraph}
\usepackage{tocloft}
\usepackage[parfill]{parskip}
\usepackage{endnotes}
\let\footnote=\endnote
\usepackage[vskip=-\parskip]{quoting}
\usepackage{setspace}
\usepackage{etoc}
\usepackage{authblk}

\usepackage[normalem]{ulem}

\begin{document}

\title{On two existing approaches to statistical analysis of social media data}
\author[1]{M. Patone\thanks{M.Patone@soton.ac.uk}}
\author[1,2]{L.-C. Zhang \thanks{L.Zhang@soton.ac.uk}}
\affil[1]{\small Department of Social Statistics and Demography, Univ. of Southampton, UK}
\affil[2]{\small Statistisk sentralbyr{\aa}, Norway}
\maketitle

\begin{abstract}
Using social media data for statistical analysis of general population faces commonly two basic obstacles: firstly, social media data are collected for different objects than the population units of interest; secondly, the relevant measures are typically not available directly but need to be extracted by algorithms or machine learning techniques. In this paper we examine and summarise two existing approaches to statistical analysis based on social media data, which can be discerned in the literature. In the first approach, analysis is applied to the social media data that are organised around the objects directly observed in the data; in the second one, a different analysis is applied to a constructed pseudo survey dataset, aimed to transform the observed social media data to a set of units from the target population. We elaborate systematically the relevant data quality frameworks, exemplify their applications, and highlight some typical challenges associated with social media data.
\end{abstract}



\section{Introduction}

There has been a notable increase of interest from researchers, companies and governments to conduct statistical analysis based on social media data collected from platforms such as Twitter or Facebook (see e.g.  \cite{kinder2014always, braojos2015small, he2013social, bright2014use, falco2018challenges}). At the same time there is also a growing concern for the quality issues associated with these new types of data \citep{boyd2012critical, bright2014use, hsieh2017total, sloan2017sage, halford2017understanding}.

The aim of this paper is to examine and summarise two existing approaches to statistical analysis based on social media data, when the analysis otherwise would have been possible based on the traditional approach of survey sampling. To fix the scope, let $U = \{ 1, 2, ..., N\}$ be a target population of \emph{persons}. Let $y_i$ be an associated value for each $i\in U$. Let the parameter of interest be a function of $y_U = \{ y_1, ..., y_N\}$, denoted by
\[
\theta = \theta(y_U)
\]
For instance, $\theta$ can be the population total or mean of the $y$-values. The quality of sample survey data can generally be examined with respect to two dimensions: representation and measurement (Groves, 2004). The representation dimension concerns the relationship between $U$ and the \emph{observed} set of persons, denoted by $s$. For example, $s$ suffers from under-coverage if there are persons in $U$ who have no chance of being included in $s$. The measurement dimension concerns the potential discrepancy between $y_i$ and the \emph{obtained} measures, denoted by $y_i^*$ for $i\in s$. For instance, $y_i^*$ may be subjected various causes of measurement error, such that $y_i^* \neq y_i$ for some persons in $s$.

Thus, to use instead social media data in this context, one needs to address two fundamental challenges with respect to each quality dimension. Firstly, social media data are initially organised around different units than persons; secondly, the relevant measures typically cannot be directly observed but need to be processed using algorithms or machine learning techniques. For example, one may like to make use of the relevant tweets to estimate the mean of a value associated with the resident population of a country. The directly observed unit (or data object) is then the tweets, whereas the statistical unit of interest is the residents. Next, instead of using designed survey instruments to measure the value of interest as one could in survey sampling, one will need to process a proxy to the target value from the Twitter texts by means of text mining.

Two existing approaches can be discerned in the literature. In what we refer to as the \emph{one-phase approach}, statistical analysis is directly applied to the observed social media data that are organised around other units than persons; whereas, in the \emph{two-phase approach}, a different analysis is applied to a constructed \emph{pseudo survey dataset}, after transforming the observed social media data to a set of persons from the target population. Thus, continuing the example above, by the one-phase approach one may conjure a function of the observed tweets as an estimator of the target population mean directly; whereas, by the two-phase approach one would try to identify and deduplicate all the tweets of the same person, and to process all the relevant Twitter texts to produce a proxy to the target $y$-value associated with that person, before applying an appropriate analysis.

In this paper we shall delineate these two approaches more generally and systematically than they have hitherto been treated in the literature, where the Social Media Index for Dutch Consumer Confidence \citep{daas2014social} serves as a typical case of the one-phase approach, and the ONS study on residency and mobility data constructed from geolocalised tweets \citep{swier2015geolocated} is used to illustrate the construction of pseudo survey dataset under the two-phase approach. We shall elaborate the relevant data quality frameworks and methodologies, and highlight some typical challenges to statistical analysis.

The rest of the paper is organised as follows. In Section \ref{general}, we systematise and describe in greater details the general issues of representation and measurement of social media data. In Section \ref{approach1} and \ref{approach2}, we delineate and examine the one-phase and two-phase approaches, respectively. Finally, some concluding remarks are provided in Section \ref{final}.

\section{General issues of representation and measurement} \label{general}

\subsection{Representation} \label{representation}

It is well recognised that social media platforms are not representative of the general population \citep{blank2017representativeness, mellon2017twitter}. For instance, in terms of demography, it has been shown that US users of Twitter and Facebook tend to be younger and more educated than the general population \citep{greenwood2016social} and they tend to live in urban areas \citep{mislove2011understanding}. Moreover, non-representative demographics tend to be confounding with other relevant attributes, e.g. politically active Italian Twitter users tend to be younger, better educated, male and left wings \citep{vaccari2015political}.

Twitter provides a typical example of online news and social networking site. Communication occurs through short messages, called \emph{tweets}; the act of sending tweets is called \emph{tweeting}. To be able to tweet, an account needs to be created. To register a user has to provide an email address, a username and a password. A user can be a person, a business, a public institution, or even softwares (bots), etc. In case of person, the user is not obliged to create an account reflecting her physical persona. Optional fields include a profile picture, a bio and a location, which are neither verified nor expected to accurately characterise the user. By default tweets are publicly available, although the user may change the privacy setting to make it private. Each tweet can be original, a reply to another tweet or a copy of a different tweet, known as a retweet. It can mention a username account ($@$) to address a specific user, and it can contain hashtag ($\#$) to declare the topic of the tweet. Hashtags offer a way to categorise tweets into specific topics (e.g. a tv show, a sport event, a news story). Some events such as football matches, film festivals or conferences may have an official hashtag under which the relevant tweets about the event is classified. Hashtags can also be user-specific and not intelligible to the general public.

As in the Twitter example, one can identify two directly observable units of data on most social media platforms, which we will refer to as the \emph{post} and the \emph{account}:
\begin{description}
\item[\emph{\textmd{Post}}] We use the generic term post to refer to the immediate packaging of social media content, which otherwise has a platform-specific name: Facebook has posts, Twitter has tweets and Instagram uses picture, etc.

\item[\emph{\textmd{Account}}] An account is the ostensible generator of a post. As in Twitter, the user(s) operating a social media account can be different entities including but not limited to persons. Moreover, the same user can have multiple accounts, but the connections between these accounts and the user are not publicly accessible.
\end{description}

Denote by $P$ and $A$, respectively, the totality of all the posts and accounts on a given social media platform. There is a many-one relationship from posts to the active accounts, denoted by $A_P = a(P)$, and the inactive accounts $A\setminus A_P$ is non-empty in general. Next, there is a many-one relationship from accounts to the users, denoted by $b(A)$. The \emph{observable} persons are given by the joint set of the target population $U$ and $u_{AP} = b(A_P) = b\big( a(P) \big)$, i.e. via the active accounts. Moreover, $U\setminus u_{AP}$ is non-empty as long as there are persons not engaged with the given social media platform, and $u_{AP} \setminus U$ is non-empty as long as they are other users than persons. These relationships are summarised in Table \ref{tab-units}.

\begin{table}[ht]
\begin{center}
\caption{Many-one relations $a$ from post to account, and $b$ from account to user} \vspace{-3mm}
\begin{tabular}{| l | c | c | c | c |} \hline
& Post & Account & Person \\ \hline
Totality & $P$ & $A$ & $U$ \\ \hline
Observable & $P$ & $A_P = a(P)$ & $U \cap u_{AP},~ u_{AP} = b(A_P) = b\big( a(P)\big)$\\
& & $A\setminus A_P \neq \emptyset$ & $U\setminus u_{AP} \neq \emptyset,~ u_{AP}\setminus U \neq \emptyset$ \\ \hline
Sample & i. $s_P \subset P$ & i. $s_A = a(s_P)$ & $U \cap s_{AP}, ~ U\setminus s_{AP} \neq \emptyset,~ s_{AP}\setminus U \neq \emptyset$ \\
 & ii. $s_P \subset a^{-1}(s_A)$ & ii. $s_A \subset A$ & i. $s_{AP} = b\big( a(s_P) \big)$, ii. $s_{AP} = b(s_A)$\\ \hline
\end{tabular} \label{tab-units} \end{center}
\end{table}

Next, a common way of collecting data from a given social platform is via the public APIs, either directly or indirectly through third-party data brokers; Web scraping provides another option, albeit with unclear legal implications at this moment. Via the APIs, a sample of posts or, less commonly, accounts is harvested directly from the social media company and the obtainable sample depends on the company's terms and conditions. Depending on the API, the obtained datasets may differ in terms of being real-time or historical, or the amount of data that is allowed for.

Take again Twitter for example. The \texttt{Streaming API} returns two possible samples: a 1\% sample of the total firehose (the firehose is the totality of tweets ever tweeted), without specifying any filter; or a sample of posts on specific keywords or other metadata associated to the post. However, if the number of posts matching these filters is greater than 1\% of the firehose, the Twitter API returns at most 1\% of the firehose. In addition, historical tweets can be retrieved using the \texttt{Search API}, which provides tweets published in the previous 7 days, with a selection based on ``relevance and not completeness'' (Twitter Inc.). For both APIs, Twitter does not provide the details of the process involved, nor guarantees that the sampling is completely random. See e.g. studies that have been conducted to understand and describe how the data generation process works with Twitter \citep{morstatter2013sample, gaffney2014data, gonzalez2014assessing, wang2015should}.

Sampling of accounts is less common, which is only feasible if the usernames are known in advance. Consider the case where the interest is on the political candidates during an election. If a complete list of their usernames are available, sampling can be performed by the analyst; all the posts generated by the sample accounts on the social media platform can possibly be retrieved. The approach is only applicable when the group is made of `elite' users (of known people), rather than `ordinary' users; for instance it is not always possible to identify all the eligible or potential voters. \cite{rebecq2015extension} uses the user ID number to randomly select a set of users from Twitter. A list of number from 1 to $N$, where $N$ represents the total of the Twitter ID numbers generated so far can be used as a frame of the Twitter accounts. However, it has been noted that some of the ID numbers are missing, allegedly because of privacy issues and that $N$ is not known.

Thus, the actually observed units are generally either a subset of $P$ or $A$ to start with. An initial observed \emph{sample} of posts, denoted by $s_P \subset P$, can lead one to a corresponding sample of accounts $s_A = a(s_P)$ and then, in principle, a sample of users $s_{AP} = b\big( a(s_P) \big)$. Given a sample $s_A$ directly selected from $A$, we can possibly acquire a sample of users $s_{AP} = b(s_A)$ and a sample of associated posts, denoted by $s_P = a^{-1}(s_A)$. The observed sample of persons are given by the joint set of $U$ and $s_{AP}$. Again, both $U\setminus s_{AP}$ and $s_{AP}\setminus U$ are non-empty in general. The relationships are summarised in Table \ref{tab-units} as well.

\subsection{Measurement} \label{measurement}

Unlike in sample surveys, social media data are not generated for the purpose of analysis. They have been referred to as  ``found data'' \citep{groves2011three, taylor2013data} to emphasise their non-designed origin. One can only decide what is best to do with the data given the state in which they are found. In light of the discussion of representation above, the obtained measures are either associated with the sample of posts or accounts. These may be based on the content of a post such as a text or an image, or the metadata of a post or account, such as the geo-location of a post or the profile of an account. In addition, one may observe the \emph{network} relationships between posts, accounts or users.

Take the Twitter for example. While Twitter does not provide the information whether a user is a parent or not, it may sometimes be possible to infer that the user behind a tweet is a parent based on its content. Similarly, while Twitter does not provide the location of a user, it is sometimes possible to infer this from the location (or content) of the relevant tweets. Finally, retweeting or the inclusion of certain hashtags may reveal certain network relationships between the different users.

With respect to the measurement of interest, according to \cite{japec2015big} and \cite{bright2014use}, social media data are seen to provide the opportunity to study the following social aspects: 1. to capture what people are thinking, 2. to analyse public sentiment and opinion, and 3. to understand demographics of a population. More generally, we shall distinguish among three types of data extraction from the sample posts and accounts:

\begin{description}
\item[\emph{\textmd{Content}}] Thought, opinion and sentiment provide typical examples of content extraction, which are the direct interest of study. Sentiment analysis is a common technique for extracting opinion-oriented information in a text. However, social media posts present some distinct challenges, because the expressions may be exaggerated or too subtle \citep{pang2008opinion}. Moreover, the posts on social media are public by nature, such that a user may easily be influenced by other opinions, or she may want to project an image of herself which does not necessarily represent the truth.

\item[\emph{\textmd{Feature}}] Demographics, location and socio-economic standing are common examples of feature extraction, when these are not the direct interest of study but may be useful or necessary for disaggregation and weighting of the results. Various techniques of `profiling' have been used for feature extraction. For instance, \cite{daas2016profiling}  and \cite{yildiz2017using} consider the problem of estimating age and gender of Twitter users based on the user's first name, bio, writing style and profile pictures. Or, \cite{swier2015geolocated} derive the likely place of residence of a user, from all the geo-located tweets that the user has posted. Completely accurate feature extraction is generally impossible regardless of the techniques.

\item[\emph{\textmd{Network}}] Directional posting, reposting, sharing, following and referencing all provide the possibility of observing network relationships among the posts, accounts or users. Common interests regarding the pattern and interaction among social network actors include  identifying the most influential actor, discovering network communities, etc. \cite{tabassum2018social} provide an overview for social network analysis.
As an example of network extraction from Twitter, it is currently possible to collect the followers' accounts of a given account ID using the API call \texttt{GET followers/ids}. For each submitted call of an account ID, the API may return a maximum of 5000 followers, with a further limit of 15 submitted calls within every 15 minutes. To obtain the complete list of followers of one account with 75000 followers would then require at least 15 minutes. The possibility and ease of network extraction is thus to a large extent limited by the APIs provided for a given social media platform.
\end{description}

In light of the above, whether by content, feature or network extraction from available social media data, one should generally consider the obtained measures as proxy values to the ideal target values. Of course, measurement errors are equally omnipresent in sample surveys. For instance, survey responses to questions of opinion may be subjected to mode effects, social desirability effects and various other causes of measurement error (e.g. \cite{biemer2011measurement}). So there is certainly scope for exploring social media data for relevant studies.

There is a noteworthy distinction between measurement errors in survey and social media data. In sample surveys, a measurement error does not affect the representation of the observed sample. The matter differs with social media data. For instance, when relevant accounts to a study are selected based on the metadata of an account, such as place of residence, errors can arise if the information recorded at the time of registration is not updated despite there has actually been a change of the situation. Such an error can then directly affect which accounts are selected for the study, i.e. the representation dimension of data quality. Similarly, one may fail to include a post in a study if it is classified as not containing the relevant opinion of interest.

It is easily envisaged that combining multiple platforms, such as Twitter and Linkedln, can be useful for enhancing the accuracy of data extraction,  although we have not been able to found any documented examples. This could be due to ethical reasons or the limitations imposed by the terms of conditions of the social media companies. An addition concern could be the `interaction' between representation and measurement just mentioned above, where e.g. the accounts for which data combination is possible are subjected to an extra step of selection from the initially observed sample of accounts.

\section{One-phase approach} \label{approach1}

In the one-phase approach, one needs to estimate the target parameter $\theta = \theta(y_U)$ directly from the obtained measures, denote by $z_j$, associated with a different observed set of units $s_P$ or $s_A$, despite the differences to $y_i$ and $U$.

To see why this may be possible at all, consider the following example. Suppose one is interested in the totality of goods ($\theta$) that have been purchased in a shop over a given time period. One could survey all the people who have been in the given shop during the period of interest and ask what they have purchased. The population $U$ then consists of all the relevant persons and $y_i$ is the number of goods they have purchased (possibly over multiple visits to the shop). Alternatively, $\theta$ can be defined based on the transactions registered over the counter. The population $P$ consists then of all the relevant transactions, and $z_j$ is the number of goods associated with each transaction $j\in P$. Clearly, despite the differences in $(y_i, U)$ and $(z_j, P)$, either approach validly aims at the same target parameter $\theta$.

Below we reexamine the Social Media Index \citep{daas2014social} as an application, to formalise this approach and the relevant quality issues and methodological challenges.

\subsection{Case: Social Media Index (SMI)}

\begin{figure}[h!]
  \centering
  \includegraphics[width=.80\textwidth]{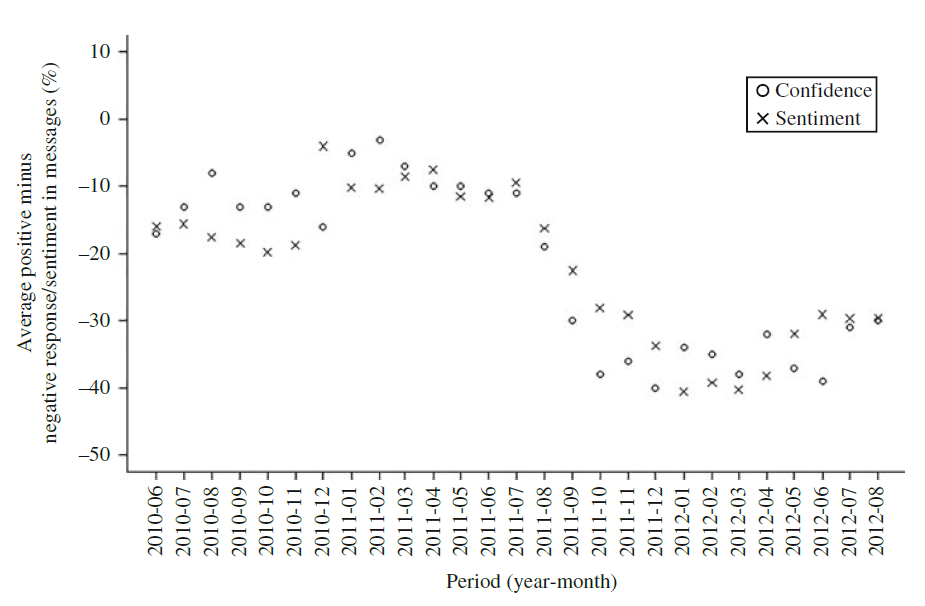}
  \caption{Comparison of Dutch CCI and SMI on a monthly basis. A correlation coefficient of $0.88$ is found for the two series \citep{daas2015big}.}\label{DaasIndex}
\end{figure}

Every month, Statistic Netherlands conducts a sample survey to compute the Consumer Confidence Index (CCI). It is based on a questionnaire of people's assessment of the country economy and their financial situation. As part of the research on the use of social media data in official statistics \citep{daas2014social, daas2015big}, the authors collected posts from different social media platforms and constructed the Social Media Index (SMI) from these posts. They observed and compared the CCI and SMI over time and concluded that the two series are highly correlated (see Figure \ref{DaasIndex}).

The SMI is constructed as an index that measures the overall sentiment of social media posts. The posts were purchased, in the time period between June 2010 and November 2013, from the Dutch company Coosto, which gather social media posts written in the Dutch language on the most popular social media of the country (Facebook, Twitter, LinkedIn, Google+ and Hyves). Coosto also assigns a sentiment classification, positive, neutral or negative to each post based on sentiment analysis \citep{pang2008opinion}, which determines the overall sentiment of the combination of words included in the text of the post. A neutral label is assigned when the text does not show apparent sentiment.

Let $P_t$ be the totality of all the observed posts in month $t$. Let $s_{P,t}$ be a subset of posts that are selected from $P_t$. Let $m_t$ be the size of $s_{P,t}$. The posts included in $s_{P,t}$ can have positive, neutral or negative sentiment value, respectively denoted by $z_j = 1, 0, -1$, for $j\in s_{P,t}$.  The SMI is calculated as the percentage difference between the positive and negative posts in $s_{P,t}$, i.e. a function of $z_{s_{P,t}} = \{ z_j; j\in s_{P,t}\}$:
\[
  \text{SMI}_t = \text{SMI}(z_{s_{P,t}}) =  \frac{100}{m_t} \sum_{j\in s_{P,t}} z_j ~.
\]

\cite{daas2014social} experimented with different ways of selecting the sample $s_{P,t}$. The choices involve a decision about which social media platforms to include, and whether to accept all the posts from an included platform or only certain groups. The groups can be filtered using a set of keywords, such as posts containing personal pronouns like `I', `me', `you' and `us', or words related to the consumer confidence or the economy, or words that are used with high frequency in the Dutch language. The idea is that selecting only certain groups of posts could effect the association between the SMI and the CCI. For instance, from a previous study \citep{daas2012twitter} the same authors found that nearly 50\% of the tweets produced in the Netherlands can be considered a `pointless bubble'. In the end $s_{P,t}$ is chosen to include all the Facebook posts and filtered Twitter posts, for which the resulting SMI achieved the highest correlation coefficient with the CCI (Figure \ref{DaasIndex}).

Finally, considering the SMI as an estimator with its own expectation and variance, let
\begin{align}\label{smi}
\text{SMI}_t = \xi_t + d_t ~,
\end{align}
where $\xi_t$ is the expectation of the SMI, and $d_t$ has mean 0 and variance $\tau_t^2$.

\subsection{Formal interpretation}

To assess the SMI as a potential replacement of the CCI, let us now formalise the CCI and its target parameter. Let $U_t$ be the Dutch \emph{household} population in month $t$, which is of the size $N_t$. Let $y_i$, for $i\in U_t$, be a consumer confidence score for household $i$ based on positive, neutral or negative responses to five survey questions. The target parameter of the CCI is given by
\[
\theta_t = \theta(y_{U_t}) = \frac{100}{N_t} \sum_{i\in U_t} y_i ~.
\]
The CCI based on the sample survey is an estimator of $\theta_t$, which can be given by
\begin{align}\label{cci}
\text{CCI}_t = \theta_t + e_t ~,
\end{align}
where $e_t$ is the sample survey error of the CCI. For our purpose here, we shall assume that $e_t \sim N(0, \sigma_t^2)$, i.e. normally distributed with mean 0 and variance $\sigma_t^2$.

Now that there is a many-one relationship between persons and households, the generic relationships from posts to persons apply equally from posts to households. The households corresponding to the SMI sample $s_{P,t}$ can thus formally be given as
\[
s_t = U_t \cap a\big( b(s_{P,t}) \big) ~.
\]
Let $s_t$ be of the size $n_t$. Let the target parameter defined for $s_t$ be given by
\[
\theta_{s,t} =  \theta(y_{s_t}) = \frac{100}{n_t} \sum_{i\in s_t} y_i ~.
\]

In order to replace the CCI by the SMI, it is now clear that one would like to have $\theta_t = \xi_t$.
However, given the underlying relationship between the social media data posts and the target population, one can only establish an analytic connection between $\xi_t$ and $\theta_{s,t}$, based on the relationship between $(z_j, s_{P,t})$ and $(y_i, s_t)$. It is therefore clear that the principal difficulty for the one-phase approach in this case is the lack of an explicit connection between $\xi_t$ and $\theta_t = \theta(y_{U_t})$, or between $\text{SMI}(z_{s_{P,t}})$ and $\theta(y_{U_t})$. Moreover, it seems that in such situations external validation will be necessary in order to establish the validity of the analysis results based on social media data, which we consider next.

\subsection{Statistical validation}

In the case of the SMI, one does have the possibility of validating its statistical relationship to the CCI, despite the lack of an analytic connection between the two. As can be seen in Figure \ref{DaasIndex}, the two indices display a high correlation with each other over time: the empirical correlation coefficient is 0.88 over the 27 months displayed. We now formulate a test to exemplify a possible venue for statistical validation in similar situations.

As a conceivable scenario in which the SMI can replace the CCI, we set up the null and alternative hypotheses below:
\[
H_0 : \theta_t - \xi_t  = \mu  \quad vs. \quad H_1 : \theta_t - \xi_t  \neq \mu ~,
\]
i.e. whether or not the target parameters of the SMI and CCI differ by a constant over time. For our purpose here, we shall make a simplifying assumption that $\tau_t^2 = 0$, and thereby remove the conceptual distinction between SMI as an estimator and its theoretical target $\xi_t$. In light of the large amount of posts in $s_{P,t}$, the assumption seems plausible. It follows then from \eqref{smi} and \eqref{cci} that, under $H_0$, we have
\[
X_t = \text{CCI}_t - \text{SMI}_t = \mu + e_t ~,
\]
where $e_t \sim N(0, \sigma_t^2)$. Thus, one may compare the total deviation of $X_t$ from its mean $\bar{X} = \sum_{t=1}^T X_t$, over the available $T$ time points, to the variances of the CCI: the larger the total deviation exceeds that which is allowed for by the CCI variances, the stronger is the evidence against $H_0$ compared to $H_1$.

Formally, let $P = I - {\bf 1} {\bf 1}^{\top}/T$, where $I$ is the $T\times T$ identity matrix and ${\bf 1}$ is the $T\times 1$ unity vector, and the matrix $P$ is idempotent such that $P P^{\top} = P P = P$. We have
\begin{align*}
& E(P X) = {\bf 0} \qquad\text{for}\quad X = (X_1, ..., X_T)^{\top} ~,\\
& V(P X) = P \Sigma P \qquad\text{for}\quad \Sigma = \mbox{Diag}(\sigma_1^2, ..., \sigma_T^2) ~.
\end{align*}
The diagonal matrix $\Sigma$ corresponds to the assumption that the CCI's are uncorrelated over time. If this is not the case, one may specify the true covariance matrix appropriately, without this affecting the generality of the following development.
Now that ${\bf 1}^{\top} PX \equiv 0$, one of the component is redundant. Let $X' = (PX)_{(-t)}$ on deleting the $t$-th component of $PX$, for any $1\leq t\leq T$. Let $Q$ be the correspond $(T-1)\times (T-1)$ sub-matrix of $P \Sigma P$, such that $X'$ has the $T-1$-variate normal distribution
\[
X' \sim N({\bf 0}, Q) ~.
\]
Let $L L^{\top} = Q$ be the Cholesky decomposition with lower-triangular $L$, such that
\[
L^{-1} Q (L^{-1})^{\top} = L^{-1} L L^{\top} (L^{-1})^{\top} = I_{(T-1)\times (T-1)}
\]
and
\[
R = L^{-1}  X' \sim N({\bf 0}, I) ~.
\]
A test statistic for $H_0$ can thus given as
\[
D = R^{\top} R \sim \chi_{T-1}^2 ~.
\]

\begin{table}[ht]
\begin{center}
\caption{Approximate values of CCI and SMI in Figure \ref{DaasIndex}} \vspace{-2mm}
\begin{tabular}[t]{|l |c |c |c |c |c |c |c |c |c |} \hline
$t$ & 1 & 2 & 3 & 4 & 5 & 6 & 7 & 8 & 9 \\
CCI & -17 & -13 & -8 & -12.5 & -12.5 & -11 & -15 & -5 & -2.5 \\
SMI & -16 & -15 & -17.5 & -17.5 & -20 & -18 & -4 & -10 & -10 \\ \hline
$t$ &10 & 11 & 12 & 13 & 14 & 15 & 16 & 17 & 18 \\
CCI  & -7 & -10 & -10 & -11 & -11 & -19 & -30 & -38 & -35.5 \\
SMI & -8 & -7.5 & -11.5 & -11.5 & -9 & -16.5 & -22.5 & -28.5 & -29.35 \\ \hline
$t$ & 19 & 20 & 21 & 22 & 23 & 24 & 25 & 26 & 27 \\
CCI & -40 & -34 & -35 & -37 & -32 & -36.5 & -39 & -30 & -29\\
SMI & -33.5 & -40.5 & -39 & -39.5 & -37 & -32 & -29 & -29.5 & -29.5\\ \hline
\end{tabular} \label{tab-index}
\end{center}
\end{table}

\begin{figure}[ht]
\begin{center}
\includegraphics[width=.6\linewidth]{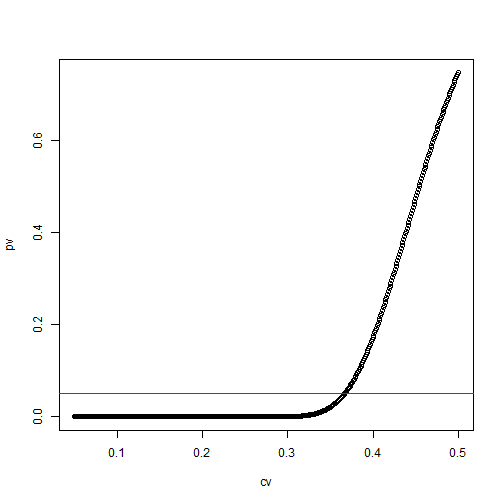}
\caption{P-values of test $H_0$ vs. $H_1$ for varying CVs, level 0.05 mark by horizontal line} \label{fig-test}
\end{center}
\end{figure}

Due to confidentiality restrictions, we are unable to obtain the actual values of the SMI and CCI in Figure \ref{DaasIndex}, nor the variances of the CCI. The calculations below serve only for the purpose of illustration. Firstly, we eyeball Figure \ref{DaasIndex} to obtain the values of the two indices approximately, which are given in Table \ref{tab-index}, where the empirical correlation coefficient between two series is 0.88 over the 27 months. Next, we stipulate the values of $\sigma_t^2$ in relation to the CCI via a constant coefficient of variation over time, denoted by $\eta$, such that $\sigma_t = \eta \text{CCI}_t$. Figure \ref{fig-test} shows the p-value of the test as $\eta$ varies from 0.05 to 0.5. The p-value exceeds 0.05 for $\eta > 0.367$. In other words, unless the CV of the CCI is larger than 36.7\%, the null hypothesis is rejected at the level of $0.05$.

\subsection{Discussion} \label{comment1}

Firstly, in the above we have considered the validity of the SMI, assuming the aim is to replace the CCI with it. Of course, even if the SMI cannot do this directly, there is still the possibility to use it to improve the CCI. \cite{van2017social} study the two indices over time using a bivariate time series model:
\begin{align*}
    	 \begin{pmatrix}
       	   Y_t\\
           Z_t
          \end{pmatrix} =
          \begin{pmatrix}
           L_t^Y \\
           L_t^Z
         \end{pmatrix} +
         \begin{pmatrix}
           S_t^Y \\
           0
         \end{pmatrix} +
         \begin{pmatrix}
           \beta^{11}\delta_t^{11} \\
           0
         \end{pmatrix} +
         \begin{pmatrix}
           \upsilon_t^Y \\
           \upsilon_t^Z
         \end{pmatrix} ~,
\end{align*}
where $Z_t$ is the SMI that is decomposed into trend $L_t^Z$ and an error term $v_t^Z$, and $Y_t$ is the CCI that is decomposed into trend $L_t^Y$, seasonal component $S_t^Y$, an error term $v_t^Y$, and $\beta^{11}\delta_t^{11}$ that is an outlier term introduced to accommodate the economic downturn at the corresponding time point. The authors find that using the SMI series as an auxiliary series slightly improves the precision of the model based estimates for the CCI, at a time when the SMI for the current month is available but not the CCI -- due to the longer production lag required for the latter. Notice that such uses of social media data as the auxiliary information for survey sampling does not pose any new theoretical challenges.

Next, disregarding the distinction between $\theta_{s,t} = \theta(y_{s_t})$ and the CCI-target $\theta_t = (y_{U_t})$, where one faces a difficulty of representation between $s_t$ and $U_t$, there is a question whether the SMI \eqref{smi} appropriately targets the `intermediary' parameter $\theta_{s,t}$. As remarked by \cite{van2017social}, the CCI survey questions involve the amount of purchases of expensive goods during the last 12 months and the tendency of households to buy expensive goods. It seems relevant to utilise internet search data and actual purchase data of such expensive goods. The implication is that one needs not to rely exclusively on social media data for content extraction, but could seek to combine them with other non-survey data. On the one hand, combining data to improve content extraction seems desirable regarding the quality of measurement.  On the other hand, doing so is likely to affect the representation dimension of data quality, as previously noticed in Section \ref{measurement}. But the quality of representation is worth examining in any case. In the current definition of SMI \eqref{smi}, each post is given the same weight. It is unclear whether this is the most appropriate treatment, because the number of posts per account or user is likely to vary in different subsets of $s_t$. Indeed, provided a method of differential weighting of the posts in $s_{P,t}$ can be justified with respect to $\theta(y_{s_t})$, targeting $\theta(y_{U_t})$ may no longer be as elusive as it is currently.

Finally, despite our focus in this paper on target parameter $\theta$ defined for $(y_i, U)$, it is conceivable that one may be interested in target parameter $\xi$ defined for $(z_j, P)$ directly. In such situations, the quality considerations are analogous to those in the case of targeting $\theta$ based on a sample $s$, for $s\subset U$, and the associated measures $y_s^* = \{ y_i^* ; i\in s\}$. A basic issue regarding representation is the fact that the sample $s_P$ is not selected from the totality $P$ according to a probability sampling design. Inference from non-probability samples have received much attention. See e.g. \cite{smith1983validity}, \cite{elliott2017inference} and \cite{zhang2018valid} for inference approaches assuming non-informative selection of the observed sample; see e.g. \cite{rubin1976inference} and \cite{pfeffermann1998parametric} for examples of approaches that explicitly adjust for the informative selection mechanism. When it comes to the measurement dimension of data quality, the traditional treatment of measurement errors in surveys \citep[e.g.][]{biemer2011measurement} may be less relevant because, as discussed in Section \ref{measurement}, content, feature or network extraction from social media data faces quite different challenges and uses quite different techniques than data collection via survey instruments.


\section{Two-phase approach} \label{approach2}
In the two-phase approach, one aims to estimate the target parameter $\theta = \theta(y_U)$ based on a pseudo survey dataset constructed from the sample of social media data to resemble a survey dataset from the target population. Denote by $s_{AP}$ the sample of statistical units in the pseudo survey dataset, and by $y_i^*$ the constructed proxy to $y_i$ for $i\in s_{AP}$.

The quality of the pseudo survey dataset $(y_i^*, s_{AP})$ with respect to the ideal census data $(y_i, U)$ can be assessed with respect to representation and measurement, under the quality framework of \cite{groves2004survey} for traditional sample survey data. The key extra concern is the necessary  transformation from the initial social media data, which is a process that does not exist for sample survey data. \cite{zhang2012topics} outlines a two-phase life-cycle model of statistical data before and during integration, respectively, which includes the transformation from multiple first-phase input datasets to the ones to be integrated at the second phase. The total-error framework of \cite{zhang2012topics} is applicable as well to the two-phase approach to statistical analysis based on social media data.

Below we examine the study of \cite{swier2015geolocated}, which aims to construct pseudo survey datasets of residence and mobility from geolocated tweets. In particular, this illustrates the generic transformation process under the two-phase approach: from the first-phase data objects (posts) to the second-phase statistical units (persons) in terms of representation, and from values obtained at the first-phase (e.g. the geolocation of a post) to the second-phase statistical variable (e.g. location of residence) in terms of measurement. Moreover, we analyse the quality of the resulting pseudo survey dataset according to the total-error framework of \cite{zhang2012topics}, and highlight some relevant methodological challenges.

\subsection{Case: Residence location from tweets}

\cite{swier2015geolocated} conducted a pilot study at the Office for National Statistics, on the potential of Twitter to provide residence and mobility data for official statistics. The main efforts concerned the construction of relevant pseudo survey datasets, which we summarise below. In addition, some simple analyses were performed, giving indications of the possible target parameters envisaged. We do not explicitly discuss these analyses here.

There were two first-phase input datasets. The first one was collected via the Twitter \texttt{Streaming API}, covering the period 11th of April to 14th of August in 2014. The search criteria involved a set of bounding rectangles covering the British Isles, for which a tailor made application was developed and deployed. Due to the fact that only a small proportion of all the tweets have precise geolocations, the obtained tweets were not affected by the 1\% threshold of the Twitter API. Nevertheless, additional terms were raised by Twitter and as a result this way of collecting data was stopped. The second dataset was subsequently purchased from GNIP (a reseller of data, now owned by Twitter), covering the period 1st to 10th of April and 15th August to 31st of October in 2014. Unlike the API data, the GNIP data was filtered by tweets with a ``GB'' country code.

Next, the two datasets were merged to create a single ``clean'' dataset. A number of tweets were removed during this process. These included e.g. the ones that were detected to be generated by Bots, or without GPS location (e.g. sent form a desktop computer), or non-GB tweets in the Twitter API data (mainly those from the Republic of Ireland), etc. In particular, mainly for privacy protection reasons, any tweet from the Twitter API was removed, if it was associated with an account outside the GNIP data.

The process of merging can therefore equally be represented as in the life-cycle model of integrated data \citep{zhang2012topics}, where linkage of separate datasets are carried out via the second-phase units associated each input datasets. In other words, one may first identify the associated Account IDs (second-phase units here) in the API and GNIP datasets, respectively; and then merge the data for the same Account ID, provided it is present in the GNIP dataset. In this case one could merge the datasets before transforming the data organised around Tweet ID to Account ID, because the two first-phase datasets share the same identifiable objects (i.e. tweets with Tweet ID)

In this way, at the beginning of the second-phase processing, one obtained a single set of GB-located tweets (81.4 million over 7 months) and the associated accounts. No further second-phase data processing took place in the representation dimension. For instance, one did not attempt to identify and classify the users behind the observed accounts. Second-phase processing in the measurement is primarily concerned with content extraction of residential location and its classification. This was carried out in the following steps.
\begin{itemize}[leftmargin=5mm]
\item The tweets associated with a given account are \emph{clustered}, using the density-based spatial clustering algorithm with noise (DBSCAN). It groups together points that are closer to each other in terms of spatial density; the cluster formed is regarded valid only if it contains a specified minimum number of points. The points in clusters below the minimum threshold are considered as noise. Of the 81.4 million tweets, 67.4 million were included in one or another cluster that contains three or more tweets. The rest clusters with only one or two tweets are classified as `invalid'.

\item Next, each valid cluster is classified as `residential', `commercial' or `others' in terms of address type, using the AddressBase that is the definitive source of address information for Great Britain. To this end, one calculates a weighted centroid of the cluster and finds the closest property to it in the AddressBase. The cluster address type is then classified according to this `nearest neighbour' property.

\item Then, for each account with one or several residential clusters, the one of them with the most tweets is classified as the `dominant' residential cluster.

\item Finally, additional classification may be attached to each cluster, such as the administrative geography it belongs to, the number of tweets it contains, the time span of these tweets (short-term if less than 31 days vs. long-term otherwise).
\end{itemize}

\subsection{Quality assessment}

Before we assess the quality of the pseudo survey dataset $(y_i^*, s_A)$ obtained under the two-phase approach when targeting $\theta$ defined for $(y_i, U)$, it is helpful to recapitulate some of the relevant technical issues, even if they do not account for all the sources of errors.

Firstly, some additional API data were actually collected on the 10th of April and 15th of August, which overlaps with the GNIP data on these two days. A small number of API tweets were found not be included in the GNIP set, all of which were associated with protected accounts -- users may opt to protect their accounts so that their tweets can only be viewed by approved followers. More generally, retrospective changes made by a user to its account or specific tweets may prevent them from being included in the historic point-in-time data available from GNIP, despite these accounts or tweets were accessible via the real-time \texttt{Streaming} API. This exemplifies a general cause for discrepancy between Twitter data collected in different ways. Two other examples of general causes are as below.
\begin{description}
\item[\emph{\textmd{Filter criteria}}] The filter criteria may not be fully compatible between the APIs and the data brokers. As explained above, in the case here, the geographic filter works differently with the \texttt{Streaming API} and GNIP.

\item[\emph{\textmd{Missing data}}] Data from APIs may be missing due to technical problems, such as moving of IT equipment or broadband router failure.
\end{description}
Next, once the data form the first phase have been merged and transformed, there are generally technical issues with data extraction and processing that are necessary at the second phase. In this case, the DBSCAN clustering of tweets is an unsupervised machine learning technique, for which it is generally difficult to verify the truthfulness of the results. The address type classification is in principle a supervised learning technique. However, it may be resource demanding to obtain a training-validation dataset, by which the classification method can be improved and its accuracy evaluated. Similarly for the classification of the dominant residual cluster.

\begin{figure}[ht]
\begin{center}
\includegraphics[width=.7\linewidth]{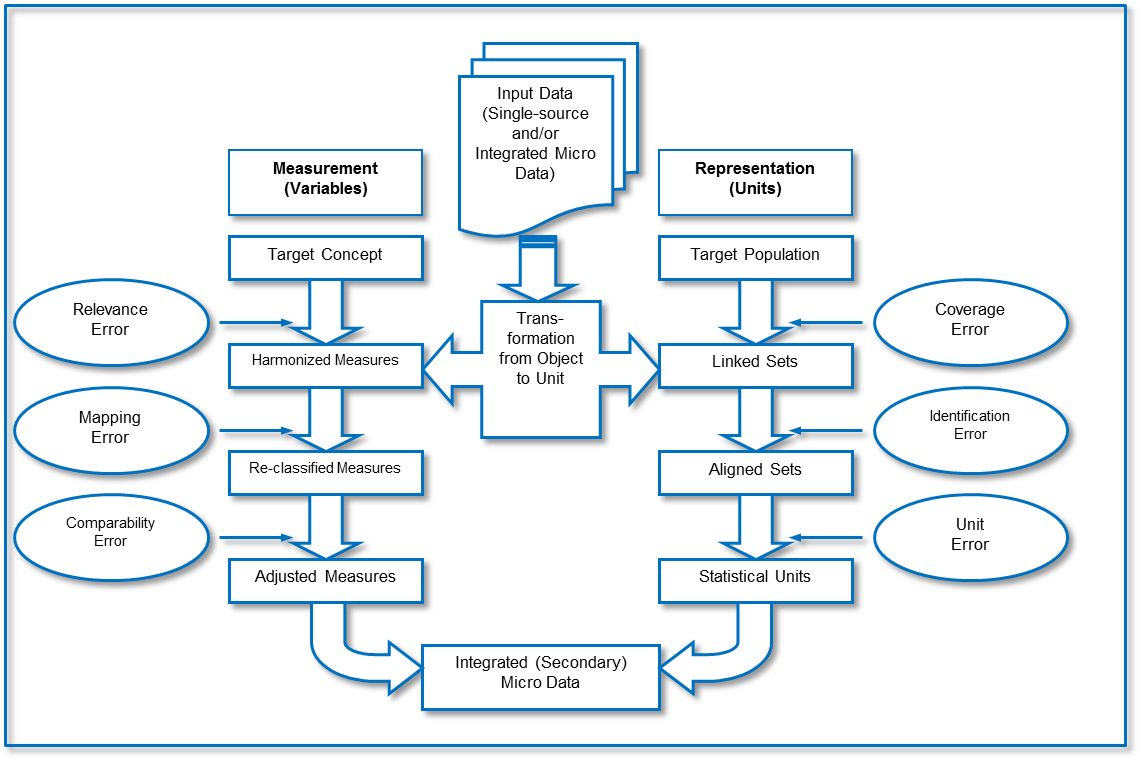}
\caption{Phase-two life-cycle model of Zhang (2012)} \label{fig-two}
\end{center}
\end{figure}

The quality of the dataset $(y_i^*, s_A)$ can be assessed according to the second-phase life-cycle model (Figure \ref{fig-two}), along the two dimensions of representation and measurement. The exact nature of the potential errors needs to be related to the envisaged analysis. Below we consider first representation and then measurement.

In terms of representation, the ``Linked Sets'' in Figure \ref{fig-two} is given by $b(s_A)$, which is subjected to coverage errors. Over-coverage is the case if $b(s_A) \setminus U \neq \emptyset$. This is unavoidable here because some of the accounts in $b(s_A)$ are not persons at all and all the Bots are not completely removed. Moreover, there may be multiple accounts in $s_A$ that correspond to the same person; such duplicates are another form of over-coverage error. Whether $s_A$ entails under-coverage depends on the assumption. For instance, let the target population $U$ be the adult residents of England. If one assumes that in principle there is an unknown but non-zero probability for everyone in $U$ to have a Twitter account and to have twittered during the 7 months in 2014, then there would be no under-coverage error of $b(s_A)$ for $U$, but only a non-probability selection issue. However, insofar as these assumptions are untenable, then there would be an under-coverage error in addition.

Next, the identification error may be an issue if domain classification of the target population needs to be based on feature extraction, which is prone to errors; whereas unit error is potentially troublesome if additional statistical units (e.g. household) need to constructed. Neither seems relevant to any of the analyses of  \cite{swier2015geolocated}.

In terms of measurement, an example of ``Harmonized Measures'' in Figure \ref{fig-two} is the dominant residential cluster here. Suppose the ``Target Concept'' is the de facto place of residence of a person. Relevance error is mostly like the case, unless everyone sends most tweets from her de facto place of residence. Or, suppose the ``Target Concept'' is whether a person is a tourist, and short-term vs. long-term classification of the dominant residential cluster is used as a proxy measure of the corresponding person. Again, relevance error is mostly like the case, unless no tourist stays longer than a month and no usual resident stops tweeting after less than a month.

Next, the mapping error is e.g. the case when someone does tweet from her de facto place of residence but the clustering-classification algorithm fails to identify it as the dominant residential cluster. This can happen e.g. if the person tweets more when at her friend's place, or if the person more often than not switches off GPS location when tweeting at home, or if the person's home is in a dense area and the chosen nearest neighbour property in the AddressBase happens to be a commercial address. Finally, the comparability error could arise if e.g. the classified dominant residential cluster is further adjusted in light of other available measures. But this was not the case in the study of  \cite{swier2015geolocated}.

In summary, the main errors of the pseudo survey dataset $(y_i^*, s_A)$ here are coverage errors in terms of representation, and relevance and mapping errors in terms of measurement.

\subsection{Discussion: Statistical analysis}

In the above we outlined the data processing required under the two-phase approach to social media data, using the study of \cite{swier2015geolocated} as the case-in-point. It is shown that the life-cycle model of \citep{zhang2012topics} can be applied as a total-error framework for evaluating the quality of the resulting pseudo survey dataset $(y_i^*, s_A)$, where $s_A = a(s_P)$. The study of \cite{swier2015geolocated} does not specify any definitive target of analysis. For a discussion of possible statistical analysis of the target parameter $\theta$ defined for $(y_j, U)$, let us consider two situations, depending on whether it involves additional datasets or not.

Consider the situation where only the pseudo survey dataset $(y_i^*, s_A)$ is to be used for an analysis targeted at $\theta(y_U)$. The first key issue regarding representation is over-coverage adjustment, from $s' = b(s_A)$ to $s = U\cap b(s_A)$, due to the fact that $s' \setminus U \neq \emptyset$. This could be either based on the mapping from $s'$ to $s$ or, provided it can be specified, from $t(y_{s'}^*)$ to $t(y_s^*)$, where $t(\cdot)$ denotes the sufficient statistics for $\theta$. Given the over-coverage adjustment, the remaining issues are non-probability representation of $s$ for $U$, and measurement discrepancy between $y_i^*$ and $y_i$ caused by lack of relevance and imperfect data extraction, similarly to what has been discussed earlier in Section \ref{comment1}.

A potentially more promising scenario is to utilise additional datasets, in order to overcome or reduce the deficiency of each dataset on its own. It is easily envisaged that integration with other Sign-of-Life data can improve the quality of the pseudo survey dataset constructed from social media data. For example, in the case of data for residence and mobility, other Sign-of-Life data on employment, education, utility services, etc. can obviously improve the classification of the dominant residential cluster, provided these data are available and can be combined with the tweets data. Making statistics based on multiple sources is a broad challenging topic. It is currently an area of active research and development. See e.g. \cite{de2017multi, di2017statistical} for overviews of related situations and methodological issues. See \cite{zhang2018use} for an overview of estimation methods in the presence of multiple proxy variables.

\section{Concluding remarks} \label{final}

In the above we systematically delineated two existing approaches to statistical analysis based on social media data. The fundamental challenge with the one-phase approach in some situations is a lack of analytic connection to the target parameter, which is defined for a different set of units and another associated measure. Nevertheless, external data can in principle be used to verify the statistical validity of this approach. Compared  to observational studies based on data subjected to non-probability selection and survey measurement errors, the key extra issues with the two-phase approach revolve around the transformation process from the initial data objects to the statistical units of interest and the algorithmic data extraction required for measurement. In addition, an explicit adjustment for the over-coverage error will be needed in many situations.

For assessment of data quality, we have demonstrated that it is possible to apply relevant total-error frameworks formulated in terms of representation and measurement of generic statistical data. In particular, for both approaches, it seems more promising if one does not simply restrict oneself to the available social media data, but seeks to combine them with additional relevant datasets, in order to overcome or reduce the deficiency of each source, despite data integration is by no means a straightforward undertaking in general.

We would like to close with a few remarks. Firstly, in the paper we have focused on target parameters that are finite-population functions. Such a parameter is often referred to as a descriptive target, in contrast to analytic target parameters that can never be directly observed, regardless how large the observed number of units and how perfect the obtained measurement may be. For example, the ordinary least squares fit of some specified linear regression coefficients based on a perfect census of the current population is a descriptive target parameter; at the same time it is an estimate of the theoretical (or super-population) values of these coefficients of the postulated regression model, i.e. the analytic target parameter in this case. Our focus on descriptive target parameters helps to simplify the exposition, since the differences between descriptive and analytic inference can be subtle and many, but are nevertheless not critical to our aim in this paper. See e.g. \cite{skinner1989analysis}, \cite{chambers2003analysis}, and \cite{skinner2017introduction} for introductions to analytic vs. descriptive inference based on sample surveys.

Next, there are certainly many similarities to statistical analysis based on administrative data. As we have demonstrated, the total-error framework \citep{zhang2012topics} for statistical data integration involving administrative sources is applicable as well to the two-phase approach based on social media data. It is worth reiterating the two extra difficulties in comparison. The first one relates to the transformation from the original data objects $P$ to the statistical units $U$. The same requirement exists equally for administrative data in general. For instance, exams are part of the initial education data objects. However, while the transformation from exams (say, $P$) to students (say, $U$) can be carried out unproblematically by the school administration, such straightforward processing is often impossible from social media data objects to the target population of interest. The second extra difficulty concerns data extraction. The available measures in the administrative sources do often suffer from relevance error. Nevertheless, the actual mapping to the ``Re-classified Measures'' (Figure \ref{fig-two}) seldom requires content or feature extraction that are necessary for social media data which, as has been discussed, is generally an additional cause of discrepancy between $y_i^*$ and $y_i$ or between $z_j$ and $y_i$.

Finally, there seems to be currently an under-explored potential regarding the rich network relationships that can be extracted from social media data. Such network relationships may be difficult to obtain via traditional survey methods, both due to the limitations of the usual survey instruments and the relatively high cognitive and memorial requirements for correct information retrieval by the respondents. In contrast, for network relationships that are directly observable on the social media platform, no subjective information processing will be needed and the errors associated with such processing are thereby avoided. Making greater use of the network relationships in social media data and developing suitable sampling and analysis methods appear fruitful venues forward, in order to harness the opportunities that have emerged with similar big data sources.

\thispagestyle{empty}
\mbox{}

\addcontentsline{toc}{section}{References}
\renewcommand{\bibname}{References}
\bibliographystyle{apalike}
\bibliography{REF}

\begin{thebibliography}{}

\bibitem[Biemer et~al., 2011]{biemer2011measurement}
Biemer, P.~P., Groves, R.~M., Lyberg, L.~E., Mathiowetz, N.~A., and Sudman, S.
  (2011).
\newblock {\em Measurement errors in surveys}, volume 173.
\newblock John Wiley \& Sons.

\bibitem[Blank and Lutz, 2017]{blank2017representativeness}
Blank, G. and Lutz, C. (2017).
\newblock Representativeness of social media in great britain: Investigating
  facebook, linkedin, twitter, pinterest, google+, and instagram.
\newblock {\em American Behavioral Scientist}, 61(7):741--756.

\bibitem[Boyd and Crawford, 2012]{boyd2012critical}
Boyd, D. and Crawford, K. (2012).
\newblock Critical questions for big data: Provocations for a cultural,
  technological, and scholarly phenomenon.
\newblock {\em Information, communication \& society}, 15(5):662--679.

\bibitem[Braojos-Gomez et~al., 2015]{braojos2015small}
Braojos-Gomez, J., Benitez-Amado, J., and Llorens-Montes, F.~J. (2015).
\newblock How do small firms learn to develop a social media competence?
\newblock {\em International Journal of Information Management},
  35(4):443--458.

\bibitem[Bright et~al., 2014]{bright2014use}
Bright, J., Margetts, H., Hale, S., and Yasseri, T. (2014).
\newblock The use of social media for research and analysis: a feasibility
  study.
\newblock {\em Report to the Department of Work and Pensions, September}.

\bibitem[Chambers and Skinner, 2003]{chambers2003analysis}
Chambers, R.~L. and Skinner, C.~J. (2003).
\newblock {\em Analysis of survey data}.
\newblock John Wiley \& Sons.

\bibitem[Daas et~al., 2012]{daas2012twitter}
Daas, P., Roos, M., Van~de Ven, M., and Neroni, J. (2012).
\newblock Twitter as a potential data source for statistics.
\newblock {\em URL http://pietdaas. nl/beta/pubs/pubs/DiscPaper\_Twitter. pdf}.

\bibitem[Daas et~al., 2016]{daas2016profiling}
Daas, P.~J., Burger, J., Le, Q., ten Bosch, O., and Puts, M.~J. (2016).
\newblock Profiling of twitter users: a big data selectivity study.
\newblock Technical report, Discussion paper 201606, Statistics Netherlands.

\bibitem[Daas and Puts, 2014]{daas2014social}
Daas, P.~J. and Puts, M.~J. (2014).
\newblock Social media sentiment and consumer confidence.
\newblock Technical report, ECB Statistics Paper.

\bibitem[Daas et~al., 2015]{daas2015big}
Daas, P.~J., Puts, M.~J., Buelens, B., and van~den Hurk, P.~A. (2015).
\newblock Big data as a source for official statistics.
\newblock {\em Journal of Official Statistics}, 31(2):249.

\bibitem[de~Waal et~al., 2017]{de2017multi}
de~Waal, T., van Delden, A., and Scholtus, S. (2017).
\newblock Multi-source statistics: Basic situations and methods.

\bibitem[di~Zio et~al., 2017]{di2017statistical}
di~Zio, M., Zhang, L.-C., and de~Waal, A. (2017).
\newblock Statistical methods for combining multiple sources of administrative
  and survey data.
\newblock {\em The Survey Statistician}, 76(July 2017):17--26.

\bibitem[Elliott et~al., 2017]{elliott2017inference}
Elliott, M.~R., Valliant, R., et~al. (2017).
\newblock Inference for nonprobability samples.
\newblock {\em Statistical Science}, 32(2):249--264.

\bibitem[Falco et~al., 2018]{falco2018challenges}
Falco, E., Kleinhans, R., and Pereira, G.~V. (2018).
\newblock Challenges to government use of social media.
\newblock In {\em Proceedings of the 19th Annual International Conference on
  Digital Government Research: Governance in the Data Age}, page 124. ACM.

\bibitem[Gaffney and Puschmann, 2014]{gaffney2014data}
Gaffney, D. and Puschmann, C. (2014).
\newblock Data collection on twitter.
\newblock {\em Twitter and society}, pages 55--67.

\bibitem[Gonz{\'a}lez-Bail{\'o}n et~al., 2014]{gonzalez2014assessing}
Gonz{\'a}lez-Bail{\'o}n, S., Wang, N., Rivero, A., Borge-Holthoefer, J., and
  Moreno, Y. (2014).
\newblock Assessing the bias in samples of large online networks.
\newblock {\em Social Networks}, 38:16--27.

\bibitem[Greenwood et~al., 2016]{greenwood2016social}
Greenwood, S., Perrin, A., and Duggan, M. (2016).
\newblock Social media update 2016. pew research center.

\bibitem[Groves, 2004]{groves2004survey}
Groves, R.~M. (2004).
\newblock {\em Survey errors and survey costs}, volume 536.
\newblock John Wiley \& Sons.

\bibitem[Groves, 2011]{groves2011three}
Groves, R.~M. (2011).
\newblock Three eras of survey research.
\newblock {\em Public Opinion Quarterly}, 75(5):861--871.

\bibitem[Halford et~al., 2017]{halford2017understanding}
Halford, S., Weal, M., Tinati, R., Carr, L., and Pope, C. (2017).
\newblock Understanding the production and circulation of social media data:
  Towards methodological principles and praxi.
\newblock {\em New Media \& Society}, page 1461444817748953.

\bibitem[He et~al., 2013]{he2013social}
He, W., Zha, S., and Li, L. (2013).
\newblock Social media competitive analysis and text mining: A case study in
  the pizza industry.
\newblock {\em International Journal of Information Management},
  33(3):464--472.

\bibitem[Hsieh and Murphy, 2017]{hsieh2017total}
Hsieh, Y.~P. and Murphy, J. (2017).
\newblock Total twitter error.
\newblock {\em Total Survey Error in Practice}, pages 23--46.

\bibitem[Japec et~al., 2015]{japec2015big}
Japec, L., Kreuter, F., Berg, M., Biemer, P., Decker, P., Lampe, C., Lane, J.,
  O’Neil, C., and Usher, A. (2015).
\newblock Big data in survey research: Aapor task force report.
\newblock {\em Public Opinion Quarterly}, 79(4):839--880.

\bibitem[Kinder-Kurlanda and Weller, 2014]{kinder2014always}
Kinder-Kurlanda, K. and Weller, K. (2014).
\newblock I always feel it must be great to be a hacker!: the role of
  interdisciplinary work in social media research.
\newblock In {\em Proceedings of the 2014 ACM conference on Web science}, pages
  91--98. ACM.

\bibitem[Mellon and Prosser, 2017]{mellon2017twitter}
Mellon, J. and Prosser, C. (2017).
\newblock Twitter and facebook are not representative of the general
  population: Political attitudes and demographics of british social media
  users.
\newblock {\em Research \& Politics}, 4(3):2053168017720008.

\bibitem[Mislove et~al., 2011]{mislove2011understanding}
Mislove, A., Lehmann, S., Ahn, Y.-Y., Onnela, J.-P., and Rosenquist, J.~N.
  (2011).
\newblock Understanding the demographics of twitter users.
\newblock {\em ICWSM}, 11(5th):25.

\bibitem[Morstatter et~al., 2013]{morstatter2013sample}
Morstatter, F., Pfeffer, J., Liu, H., and Carley, K.~M. (2013).
\newblock Is the sample good enough? comparing data from twitter's streaming
  api with twitter's firehose.
\newblock In {\em ICWSM}.

\bibitem[Pang et~al., 2008]{pang2008opinion}
Pang, B., Lee, L., et~al. (2008).
\newblock Opinion mining and sentiment analysis.
\newblock {\em Foundations and Trends{\textregistered} in Information
  Retrieval}, 2(1--2):1--135.

\bibitem[Pfeffermann et~al., 1998]{pfeffermann1998parametric}
Pfeffermann, D., Krieger, A.~M., and Rinott, Y. (1998).
\newblock Parametric distributions of complex survey data under informative
  probability sampling.
\newblock {\em Statistica Sinica}, pages 1087--1114.

\bibitem[Rebecq, 2015]{rebecq2015extension}
Rebecq, A. (2015).
\newblock Extension sampling designs for big networks.
\newblock In {\em CMStatistics 2015}.

\bibitem[Rubin, 1976]{rubin1976inference}
Rubin, D.~B. (1976).
\newblock Inference and missing data.
\newblock {\em Biometrika}, 63(3):581--592.

\bibitem[Skinner et~al., 2017]{skinner2017introduction}
Skinner, C., Wakefield, J., et~al. (2017).
\newblock Introduction to the design and analysis of complex survey data.
\newblock {\em Statistical Science}, 32(2):165--175.

\bibitem[Skinner et~al., 1989]{skinner1989analysis}
Skinner, C.~J., Holt, D., and Smith, T. M.~F. (1989).
\newblock {\em Analysis of complex surveys}.

\bibitem[Sloan and Quan-Haase, 2017]{sloan2017sage}
Sloan, L. and Quan-Haase, A. (2017).
\newblock {\em The SAGE handbook of social media research methods}.
\newblock Sage.

\bibitem[Smith, 1983]{smith1983validity}
Smith, T. (1983).
\newblock On the validity of inferences from non-random sample.
\newblock {\em Journal of the Royal Statistical Society. Series A (General)},
  pages 394--403.

\bibitem[Swier et~al., 2015]{swier2015geolocated}
Swier, N., Komarniczky, B., and Clapperton, B. (2015).
\newblock Using geolocated twitter traces to infer residence and mobility.
\newblock {\em GSS Methodology Series}, 41.

\bibitem[Tabassum et~al., 2018]{tabassum2018social}
Tabassum, S., Pereira, F.~S., Fernandes, S., and Gama, J. (2018).
\newblock Social network analysis: An overview.
\newblock {\em Wiley Interdisciplinary Reviews: Data Mining and Knowledge
  Discovery}, 8(5):e1256.

\bibitem[Taylor, 2013]{taylor2013data}
Taylor, S.~J. (2013).
\newblock Real scientists make their own data.
\newblock {\em Sean J. Taylor Blog}.

\bibitem[Vaccari et~al., 2015]{vaccari2015political}
Vaccari, C., Valeriani, A., Barber{\'a}, P., Bonneau, R., Jost, J.~T., Nagler,
  J., and Tucker, J.~A. (2015).
\newblock Political expression and action on social media: Exploring the
  relationship between lower-and higher-threshold political activities among
  twitter users in italy.
\newblock {\em Journal of Computer-Mediated Communication}, 20(2):221--239.

\bibitem[Van~den Brakel et~al., 2017]{van2017social}
Van~den Brakel, J., S{\"o}hler, E., Daas, P., and Buelens, B. (2017).
\newblock Social media as a data source for official statistics; the dutch
  consumer confidence index.
\newblock {\em Survey Methodology}, 43(2).

\bibitem[Wang et~al., 2015]{wang2015should}
Wang, Y., Callan, J., and Zheng, B. (2015).
\newblock Should we use the sample? analyzing datasets sampled from twitter’s
  stream api.
\newblock {\em ACM Transactions on the Web (TWEB)}, 9(3):13.

\bibitem[Yildiz et~al., 2017]{yildiz2017using}
Yildiz, D., Munson, J., Vitali, A., Tinati, R., Holland, J., et~al. (2017).
\newblock Using twitter data for demographic research.
\newblock {\em Demographic Research}, 37:1477--1514.

\bibitem[Zhang, 2012]{zhang2012topics}
Zhang, L.-C. (2012).
\newblock Topics of statistical theory for register-based statistics and data
  integration.
\newblock {\em Statistica Neerlandica}, 66(1):41--63.

\bibitem[Zhang, 2018a]{zhang2018use}
Zhang, L.-C. (2018a).
\newblock On the use of proxy variables in combining register and survey data.
\newblock {\em Administrative Records for Survey Methodology}.

\bibitem[Zhang, 2018b]{zhang2018valid}
Zhang, L.-C. (2018b).
\newblock On valid descriptive inference from non-probability sample.
\newblock {\em arXiv preprint arXiv:1810.00579}.

\end{thebibliography}

\end{document}